# Title: Spanning the Scales of Granular Materials: Microscopic Force Imaging

Authors: Nicolas Brodu[1,2], Joshua A. Dijksman[1,3], Robert P. Behringer[1*]

**Affiliations:**

[1]Duke University, Department of Physics, Physics Building, Science Drive, Box 90305, Durham, NC, 27708, USA.

[2]INRIA, 200 avenue de la Vieille Tour, 33405 Talence, France.

[3]Department of Physical Chemistry and Colloid Science, Wageningen University, PO Box 8038, 6700EK Wageningen, The Netherlands.

*Correspondence to: bob@phy.duke.edu

**Abstract**: If you walk on sand, it supports your weight. How do the disordered forces between particles in sand organize, to keep you from sinking? This simple question is surprisingly difficult to answer experimentally: measuring forces in three dimensions, between deeply buried grains, is challenging. We describe here experiments in which we have succeeded in measuring forces inside a granular packing subject to controlled deformations. We connect the measured micro-scale forces to the macro-scale packing force response with an averaging, mean field calculation. This calculation explains how the combination of packing structure and contact deformations produce the unexpected mechanical response of the packing, and reveals a surprising microscopic particle deformation enhancement mechanism.

**One Sentence Summary:** Microscopic three dimensional force measurements inside a dense granular packing explain the macroscopic mechanical response of granular packings under compression.

**Main Text:**

The way that a disordered packing of macroscopic particles responds to mechanical deformations, such as compression, has been investigated as early as the 17$^{th}$ century, when Stephen Hales studied the expansion of dried peas submerged in water [1]. The motivation for such studies is clear. Packings of particles surround us: coffee beans and rice; soils, embankments, many industrial processes, and geophysical processes from earthquakes to landslides involve granular materials [2], hence, the gamut of recent investigations [3, 4, 5]. Part of the relevant physics also applies to emulsion and foams, which consist of highly deformable frictionless particles, and to colloidal systems [6, 7], where thermal agitation is relevant. A major challenge for granular materials is relating microscopic contact and force details to global system response, such as applied stresses or strains. The link from micro- to macro behavior informs the mathematical link, known as a constitutive relation, between macroscopic strain or stress, and macroscopic response. These relations appear routinely for conventional materials such as Newtonian fluids (shear stress proportional to shear strain rate) or linear elastic solids (stress proportional to strain). For granular materials, there are continuum models [8], such as those used in the soil mechanics communities. Although these may be widely used, they suffer from problems: the connection between grain-scale and macro-scale behavior is still an open question, and these models are known to contain complex mathematical instabilities [9]. The lack of a well-established constitutive relation for granular materials is highly limiting; it is essential in applications where a continuum description is the only computationally feasible approach.

In order to develop models such as constitutive relations, it is essential to have direct access to microscopic experimental information that shows how granular materials respond to applied stresses or strains. The challenge is to have access to all microscopic quantities inside a granular packing, including at every contact between grains, not just for a single packing under fixed boundary conditions, but during the entire evolution of a packing under realistic deformations. Experimental probes of granular microscopic properties which can provide such structure and contact forces in three dimensional (3D) packings are limited. Photoelastic methods provide this type of information in two dimensions [10], but are hard to implement in 3D. Pioneering measurements by Brujić et al. [12] and by Zhang et al. [13] for emulsions yielded the distribution of forces at contacts, P($f$), and information on contact networks. However, they did not complete the connection between microscopic properties, such as inter-particle forces and macroscopic properties, such as the stress tensor. Tomographic techniques (such as microCT [11, 12] and confocal microscopy [13, 14]) have been successfully implemented, but they are limited in their ability to yield microscopic structure, particularly contact forces, over many closely spaced macroscopic state changes, and with a precise control of applied strains. To analyze CT data, inverse models [15, 16] of particle deformations find a possible solution, but not necessarily the actual solution. While this is reasonable, it does not yield direct measurements of the true contact forces.

In this work, we use refractive index matching tomography to provide full access to microstructure, for packings of deformable hydrogel particles that are compressed and decompressed uniaxially. Recent work using laser-scanning techniques has yielded particle locations and contact numbers [17,18]. Our work crucially adds the ability to measure contact forces in vectorial detail while straining the sample, enabling us to track the system-scale stress tensor properties over small strain steps. This feature gives access to the complete micro-macro range of mechanical details of the packing, including ingredients for constitutive modeling and its underlying physical mechanisms.

In a typical experiment, sketched in Fig.1a, 514 hydrogel particles that have been saturated with fluorescent dye are contained in a Plexiglas box. The particles are roughly, but not perfectly, spherical, with a mean diameter of 2.1 cm. The box is filled with a solution of water and polyvinylpyrrolidone (PVP) such that the index of refraction of the particles is well matched to the solution. This allows optical access to a vertical laser sheet that is scanned horizontally. A camera, whose image plane is parallel to the laser sheet, records a series of images as the laser sheet is swept. In order to reconstruct particles and contacts, we have developed dedicated tomography algorithms that yield the force (including direction) for each individual particle contact. We discuss this method further, along with additional experimental information, in the Supplementary Materials. The particles are contained in a box with five rigid transparent walls. The sixth and topmost wall is a porous piston which moves in the vertical direction, providing compression/decompression. A force gauge, which is in-line with the piston, measures the vertical force acting on the top layer of particles, and hence the pressure at the top boundary. The piston location and the force gauge give us the macroscopic strain and stress imposed on the system. For each scanned sequence of images, we infer the particle shapes, locations and all the contact vector forces, which gives us the full microstructure (Fig.1b). For the present system, friction is negligible, and forces are nearly normal to contacts. The density of the particles is also nearly matched to the fluid, so that the particles experience an effective gravity of about 0.01g (g is the acceleration of gravity).

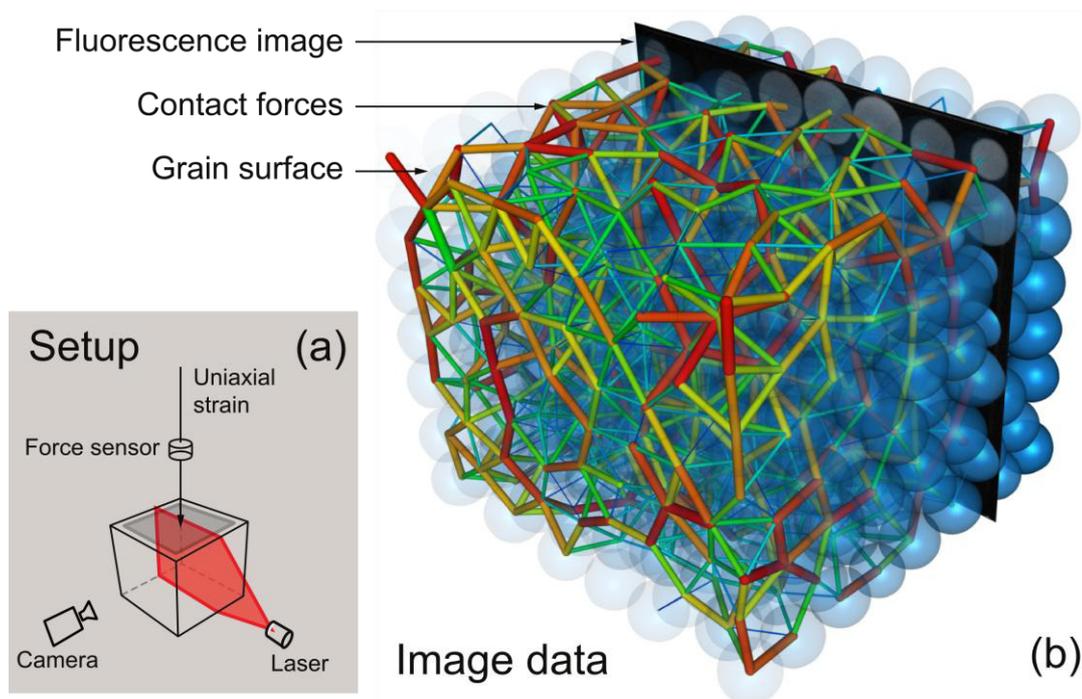

**Fig. 1**. (a) Schematic of the setup: hydrogel grains saturated with a fluorescent dye are immersed in an index-matched bath, which is subject to a uniaxial strain. A moving laser sheet illuminates cross-sections, which are imaged by a digital camera; a force sensor measures the compressive stress. (b) Superimposed image data showing part of a single slice fluorescence image (black with grey particle cross sections), the reconstructed particles (blue), and force networks (colored rods). The blue semi-transparent spherical shapes show the reconstructed particle surfaces from one complete set of slices. Rods show forces between grains, as determined from the reconstructed grain surfaces and contact deformations. Rod thickness (thin-thick) and color (blue-red) represent strength of the force (low-high) at the contact, scaled with the cumulative distribution function.

The experimental protocol consists of a series of 20 uniaxial compressions, each followed by a corresponding expansion to the original boundary configuration. Each cycle consists of a compression phase imposing a strain up to 13.4% of the initial height, followed by a decompression phase returning the top plate to its original position. A full cycle is carried out in 60 quasi-static steps of 1mm each, but the top plate does not touch the grains in the first five and last five steps. After each quasi-static step, we carry out a complete volume scan.

In order to obtain the contact forces from the reconstructed scanned images, we implement the following steps, which we discuss further in the Supplementary Materials. First, we identify the boundary of each particle, which is not perfectly spherical, including regions of contact with other particles. Then, from the areas of contact, we use linear elasticity, and the independently measured Young modulus of the particles ($E \approx 23$kPa), to infer the contact normal forces. The particles are nearly incompressible (volume change from uncompressed to fully compressed is less than 1%) so we assume that they have a Poisson ratio of $\approx 0.5$. Our best estimate of the hydrogel friction coefficient is $\mu \approx 0.03$. Hence, for these experiments, the tangential (frictional) forces are at most 3% of the corresponding normal forces, and lie below experimental resolution (the minimum average contact force is $\langle f \rangle = 10^{-2}$ N, see Supplementary Fig.S3b). We then compute the coarse-grained continuum stress tensor [19]. We integrate the normal contribution of the stress tensor over the upper boundary to obtain the force on the top plate. We emphasize that this measurement uses only information from the microscale and the independently measured particle Young's modulus. The force data resulting from the stress integration and from the independent in-line force gauge are nearly identical, as we show in Fig.2, for the last 15 of the 20 uniaxial compression/decompression cycles. Here, blue and green distinguish the macroscopic force measurement of the gauge from the microscopically derived measurement respectively. Although there is a small offset between the two different force measurements, the overall agreement is very good, and the results are very reproducible from cycle to cycle. The difference in the two types of measurements may be due to friction, which we neglected, as well as small differences in the properties of the hydrogels between compression and decompression [18]. In the remainder of this paper, we use the force values derived from the images, for consistency with other microstructure measures.

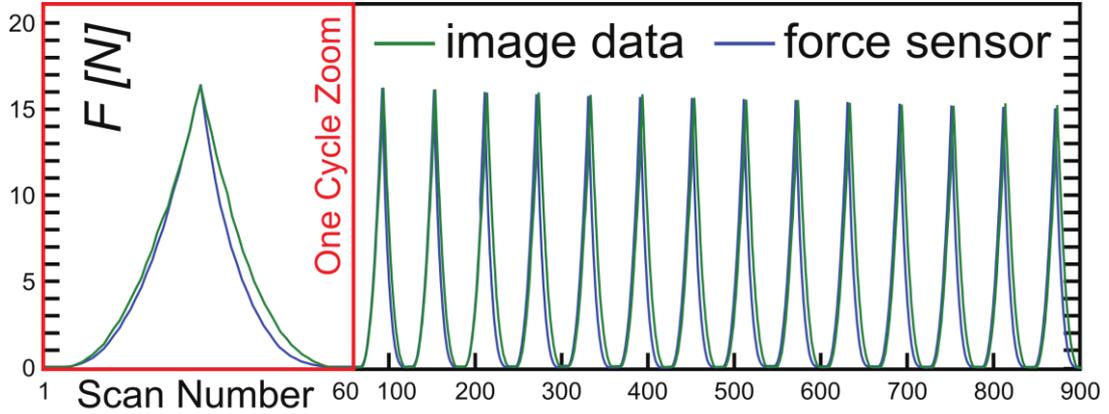

**Fig. 2**. Comparison of the compressive force *F* measured by the in-line force gauge (blue) with the force inferred from the tomographic reconstruction (green), for each of the 900 full scans taken during 15 loading/unloading cycles of the uniaxial compression experiment (the first cycle is blown-up). The top plate does not touch the grains in the first five and last five scans in each cycle, but in between, the scan numbers can be viewed as an "unfolded" measure of strain expressed in mm.

With large data sets of particle-scale data, it is possible to obtain reliable statistics on all quantities of interest. We measured data from the 20 compression cycles, but we avoid the first five cycles in the analysis as they show weak transient effects. We compute statistics (microscopic averages noted by $\langle \cdot \rangle$) only on grains which do not touch the walls, and on contacts between such grains, in order to reduce boundary effects. We also separate the compression and the decompression motion at the same compression level in order to display the full mechanical trajectory. The measures presented in Fig. 3 are typically averaged over 10,400 samples for contacts and 4,100 samples for grains.

These data allow us to link microscopic quantities to similar global properties of the packing, which we discuss in the context of Fig.3. In particular, we provide insight into the nominally power-law relation between the macro-scale force $F$ and the total strain $\Delta$: $F \propto \Delta^{\beta}$. We contrast $F$ and $\Delta$ to their microscopic equivalents, $\langle f \rangle$ the average force at contacts, and $\langle \delta \rangle$ the average contact deformation. These last two variables are related by Hertz' law, together with the radius of curvature, $r$, at contacts. A key observation is that $F(\Delta)$ and $\langle f \rangle(\langle \delta \rangle)$ follow substantially different functional relations [20].

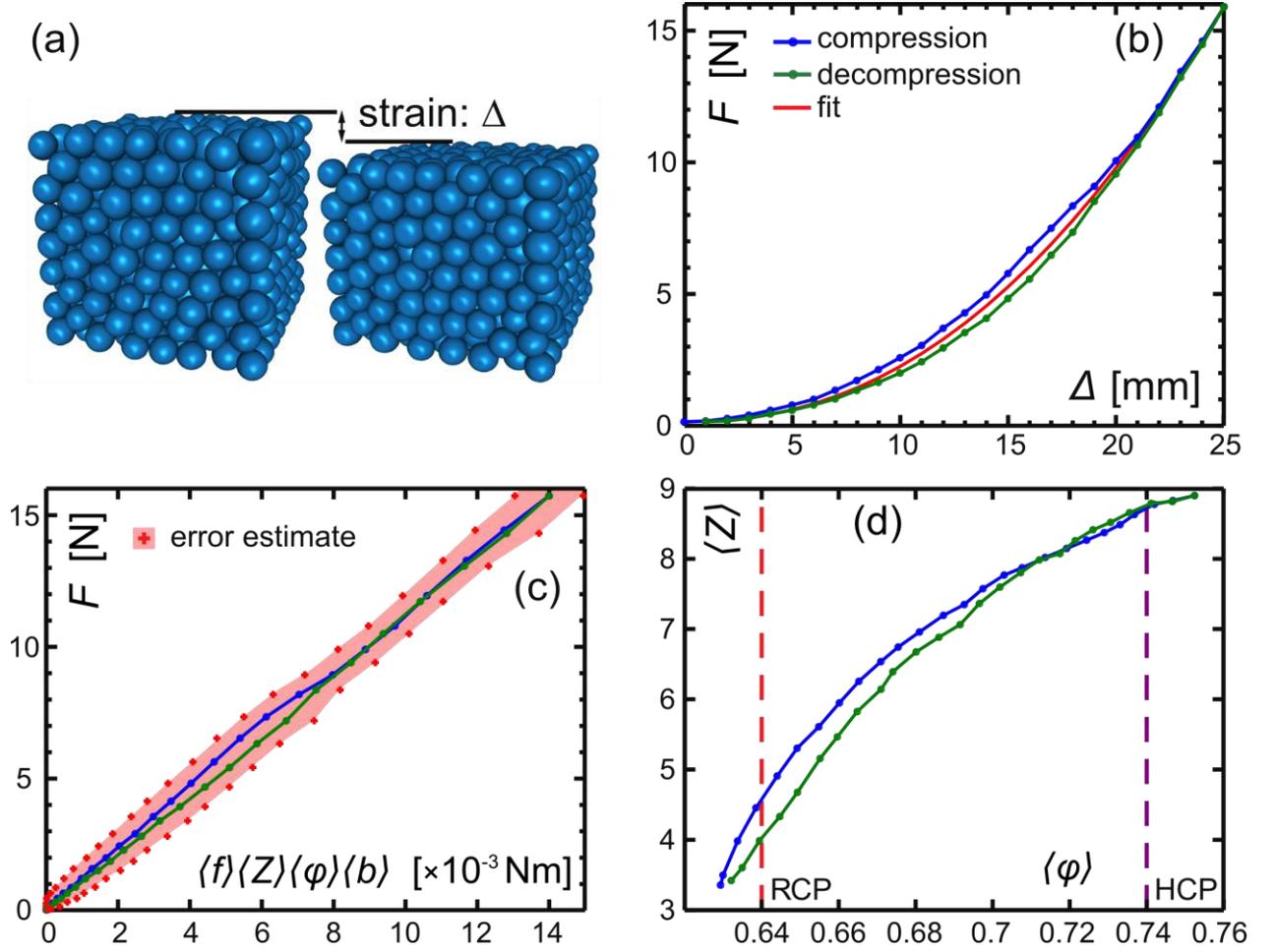

**Fig. 3.** Relations between microstructural measures and global quantities. (a) Packing reconstruction of uncompressed and fully compressed particles, indicating the global compression, (b) Empirical power law fit of the compressive force, $F$, vs. $\Delta$; the average best fit gives an exponent $\beta=2.2$. The legend applies to all panels. (c) Scaling relation, based on a mean field pressure argument, derived in Supplementary Materials. Error estimates are overly conservative. (d) The number of contacts per grain, $\langle Z \rangle$, varies with the volume fraction within the packing. Errors in $\langle Z \rangle$ are largest near jamming, where we underestimate this quantity. The random close packing (RCP) and hexagonal close packed limits (HCP) are indicated with dashed lines

For reference, Fig.3a shows the uncompressed and fully compressed states corresponding to one of the cycles. In Fig.3b, we show data for the global force, $F$, vs. the global compression $\Delta$. A fit of these data to a power law yields an exponent $\beta=2.2$, which is significantly larger than the microscopic (Hertzian) force law exponent of 1.5 measured for inter-grain forces. The key point is that $F$ depends on $\langle f \rangle$ as well as on additional microscopic properties. The pressure, exerted on the top plate, can be expressed as a combination of the averaged microscopic quantities. Using a mean field argument detailed in Supplementary Materials MM3, we obtain:

$$F \propto \langle f \rangle \langle Z \rangle \langle \varphi \rangle \langle b \rangle \qquad (\text{Eq.1})$$

where $\langle b \rangle$ is the average distance between grain centers and $\langle \varphi \rangle$ is the average packing fraction of grains within the packing (not touching the boundaries). In Fig.3c, we show this mean field relation is well satisfied. This result gives a simple explanation of why the global force response with strain has a larger effective exponent than the particle-scale force law.

Fig.3d shows data for ⟨Z⟩ vs. ⟨φ⟩. Near jamming, for ⟨φ⟩ near 0.64, we expect ⟨Z⟩ ≈ 6, whereas we measure Z between 4 and 5. This error is due to experimental uncertainty in distinguishing between two particles which are very close, but not in contact, vs. actually in contact. By contrast, near maximal compression, the experimental error in ⟨Z⟩ is much lower. For the largest packing fractions, ⟨φ⟩ > 0.74, our particles organize into a crystalline lattice (supplementary Fig. S2). Note that ⟨φ⟩ = 0.74 corresponds to the packing fraction for an ordered close-packed lattice of hard spheres (HCP), although our particles are deformable. Using an overestimate for errors in all relevant quantities, including ⟨Z⟩, we obtain the error bars indicated by crosses for the scaling relation of Fig.3c and Eq.1. Note that the scaling relation of Eq.1 is not strongly affected in absolute terms by errors in ⟨Z⟩, since the region where ⟨Z⟩ is most uncertain corresponds to small values of ⟨f⟩.

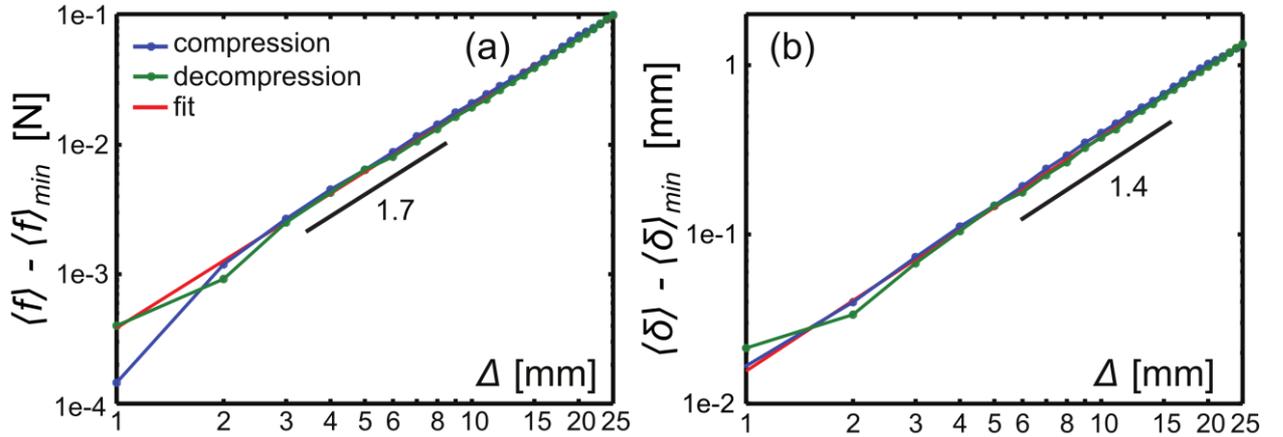

**Fig. 4**. (a) Approximate power-law relations between ⟨f⟩ and Δ, and (b) between ⟨δ⟩ and Δ. The legend applies to both panels. The fits are power laws with exponents as indicated.

Based on Eq.1, we conclude that the constitutive relation $P(\Delta)$ which yields $F \propto \Delta^\beta$ is approximately, but not strictly a power-law. For example, ⟨φ⟩, in Eq.1, varies with Δ as ⟨φ⟩ ∝ $1/(c-\Delta)$, where the constant c depends on the packing geometry. From Eq.1 we further see that F depends on factors such as the topology of the microstructure, which evolve with Δ. However, the fact that F is roughly a power-law in Δ suggests that there may be other power-law-like relations between microscopic and macroscopic properties. For instance, we find that ⟨f⟩ ∝ $\Delta^\gamma$, with an empirical exponent γ ≈ 1.7 (Fig.4a), and that ⟨δ⟩ is well approximated by ⟨δ⟩ ∝ $\Delta^\alpha$, with α ≈ *1.4*, providing we shift ⟨δ⟩ by a small offset, ⟨δ⟩$_{min}$, which is below the experimental resolution (Fig.4b). This yields the surprising finding that, with α>1, the global strain Δ produces nonlinearly amplified deformations ⟨δ⟩ at the grain level. We stress that ⟨δ⟩ appears both in Hertz' law and as part of ⟨b⟩ in Eq.1 (see MM2). This makes the nonlinear deformation amplification highly relevant, while also showing the deep entanglement of force and structure in this kind of constitutive modeling, linking all the empirical exponents $\alpha$, $\beta$ and $\gamma$ in these disordered packings.

**Conclusion**

We have presented a novel experimental technique which yields all microscopic contact and force vector information from a three dimensional packing of particles subject to controlled deformation. With these microscopic data, we have been able to verify a quantitative relation between the macroscopic mechanical response of the packing and the microscopic structural metrics of the packing. Furthermore, our experiment has revealed a surprising nonlinear enhancement of contact deformation in response to a global packing deformation. Our results have important repercussions for the understanding and modeling of granular materials. In addition, our new experimental approach has great potential for yielding more understanding of granular systems, including packing response to shear, the jamming transition, particle diffusion, effects of particle shape, etc. It supplements other 3D imaging methods, such as microCT, NMR and confocal imaging, opening up a wide range of opportunities to shed new light on the poorly understood mechanics of disordered materials.

**Acknowledgments:** This work has been supported by NASA grant NNX10AU01G, NSF grants DMR1206351, DMS1248071, and ARO grant W911NF-1-11-0110. Data is available on request from R. Behringer, bob@phy.duke.edu.

**Supplementary Materials:**

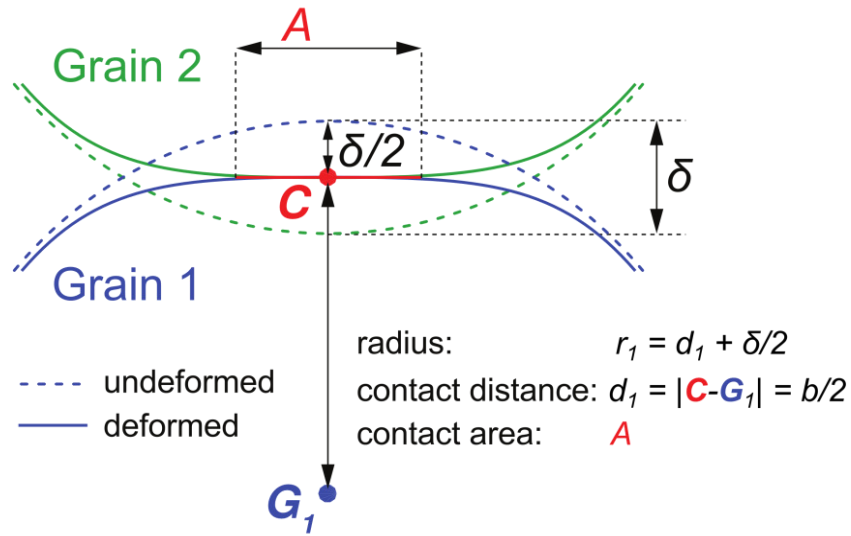

**Fig. S1**. Contact and grain geometries, sketched in a 2D plane orthogonal to the contact area. The surface of each grain (blue/green) is shown as solid lines, the hypothetical "undeformed" surface of the grains is shown as a dashed line. The deformation $\delta$ for this contact is the unknown quantity that we infer from linear elasticity. We measure the area of contact $A$, the position of the contact centroid $C$ and its distance $d$ to the center of mass $G$ of each grain. We also indicate how $r$, $d$ and $b$ are related.

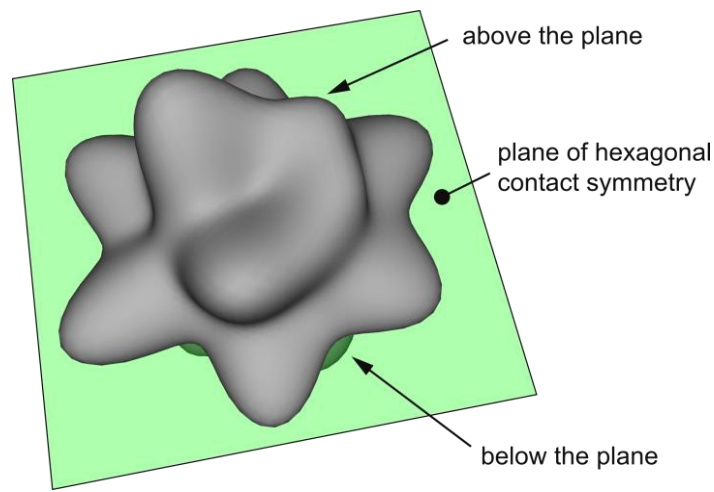

**Fig. S2**. Three-dimensional realization of the distribution of directions for contacts at the maximal compression level, $\varDelta$=25mm. Contacts directions show a median plane with an hexagonal symmetry, typical of closed packed lattices. The directions above and below that plane correspond to either face-centered cubic or hexagonal close packed variants.

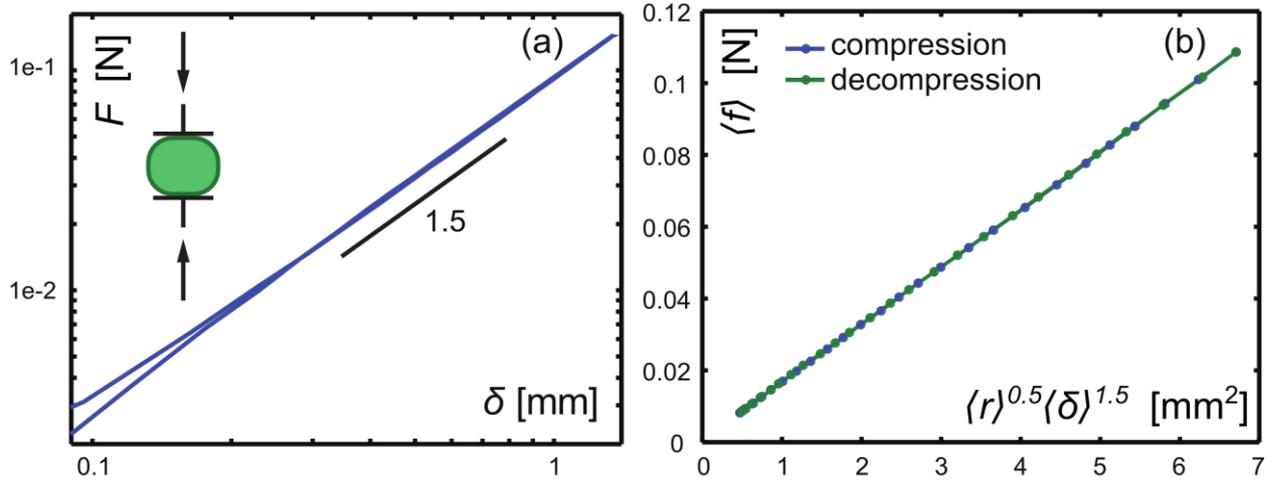

**Fig. S3**. (a) Force measured when compressing a single hydrogel particle in a microstrain analyzer. We applied a cycle of compression/decompression up to 7.2%. During the test, the grain was immersed in a bath with temperature 21°C ± 1°C. Temperature variations between 14 and 21°C do not affect the power law scaling, only the modulus. This test verifies the Hertz law, $f \propto \delta^{3/2}$, and yields the Young's modulus $E = 23$kPa ± 1kPa for this temperature. (b) Data averaged over all contacts for the 15 compression cycles described in the main text. Here, $f$ is the force at a contact, and $\delta$ is the deformation of one of the particles at the contact. Hertz law, $f \propto r^{0.5}\delta^{1.5}$ for an individual contact, also holds for averaged quantities. That is, by a similar argument to MM3 below, $\langle f \rangle \propto \langle r \rangle^{0.5} \langle \delta \rangle^{1.5}$ can be used in place of $\langle f \rangle \propto \langle r^{0.5}\delta^{1.5} \rangle$. This figure also gives a conservative estimate on our experimental force resolution of $\langle f \rangle = 10^{-2}$N in the least compressed case.

## Materials and Methods

### MM1: Tomography: inference of 3D geometrical properties from images

For a given state of our system, we measure images of the packing cross-sections. From these we infer the geometric properties of the grains and their contacts. We use standard image denoising, local gray level renormalization, and image rescaling to compensate varying optical path lengths. We combine all the slice images into a 3D representation in terms of voxels. We detect border voxels by using a combination of statistical tests and signal processing techniques to account for physical properties of the hydrogels (e.g. homogenous gray levels within grains, compensation of artifacts induced by index mismatch at grain boundaries due to their surface properties, etc). These methods are detailed in the publicly available source code. They ensure that very few false border outliers remain. We then uniquely attribute border voxels to grains: a local tangent plane is fit at each voxel, such that the plane normal determines the direction toward a rough estimate of the grain center at an average radius distance. These proto-centers are considered to be distributed around a unique grain identifier, which we compute with 3D kernel density estimation. This process yields a cloud of border voxels uniquely attributed to each grain. We then fit an analytic shape to these borders such that for each direction ***u*** (unit vector), the surface of the grain from the center of mass in that direction is given by a function $s(\boldsymbol{u})$ that best interpolates through all border voxels. This function $s$ is expressed in terms of a basis of spline functions on $S^2 \equiv \{u\}$, the unit sphere. Unlike their Cartesian counterpart (e.g. non-uniform rational B-splines (NURBS)), the triangular spherical b-splines [21] do not introduce any singularities on the sphere, and they are maximally isotropic. We use the convexity of the hydrogel grains as a regularizing condition, and we then perform a Levenberg-Marquardt least-square optimization to find the coefficients of $s$ in the spline basis, so as to best fit border voxels, discarding any remaining outliers. Once we have analytic descriptions of the grain surfaces, it is easy to compute their centers of mass and any other geometric quantity, including the areas of contact.

**MM2: Derivation of the contact force from geometric properties of the grains**

We compute the deformation, $\delta$, at a given contact from the geometric properties of the grains involved in that contact. All grains have the same Young modulus, so $\delta$ is split equally between the two grains [22]. The radius of curvature, $r$, of the undeformed surface (see Fig S1), which is different for each grain and each contact, is therefore $r = d + \frac{1}{2}\delta$. We observe that the force is given by $F = E_e r_e^{1/2} \delta^{3/2}$ and $F = E_e \delta a$, with $E_e$ the effective Young modulus, $1/r_e = 1/r_1 + 1/r_2$ the effective contact radius and $a = \sqrt{(A/\pi)}$ the radius of the area of contact. Equating the two expressions for the contact force we obtain a cubic relation which gives us $\delta$. We have independently measured the Young's modulus for our particles: $E \approx 23$ kPa, see Fig. S3a.

**MM3: Derivation of the scaling relation between F and the microscopic quantities**

The stress tensor can be computed [23] with a relation of the form $\sigma = 1/V \Sigma_{c \in V} b_c \otimes f_c$, with $V$ an averaging volume and $c \in V$ the contacts in that volume ($V$ is actually replaced by a smoothing kernel [19]). For each contact $c$, $b_c$ is the vector between the centers of the grains and $f_c$ is the force vector along the contact normal. Neglecting friction and non-sphericity, $b_c$ and $f_c$ are nearly aligned; hence, the trace tr$(b_c \otimes f_c) \approx b_c \cdot f_c$. The number of terms in the sum depends on the density of contacts, which is about $\frac{1}{2} Z \varphi$, with $Z$ the number of contacts per grain and $\varphi$ the grain volume fraction within the packing (for grains not touching the boundaries). The grains are nearly density matched with the surrounding fluid (density difference of $\approx 10$ kg/m³) so we can ignore the hydrostatic pressure. With roughly spherical grains, the isotropic pressure is very nearly proportional to tr $b_c \otimes f_c$. Hence, $p \propto b f Z \varphi$. If we write any one of the quantities, $q$, on the right as $q = \langle q \rangle + \varepsilon$, the contact force, $f$, has qualitatively different statistical properties compared to $b$, $Z$, and $\varphi$. The contact forces are distributed broadly over $0 \leq f \leq f_{max}$, while the others have relatively narrow fluctuations around a non-zero mean value. For instance, $b$ is typically twice the typical radius minus a typical small deformation $\delta$. Since $f$ is the only quantity that can change by orders of magnitude in the experiment, we use the mean values of $b$, $Z$, and $\varphi$ in the expression for $p$ and assume that deviations from the mean are uncorrelated, which means that $p \propto \langle b \rangle \langle Z \rangle \langle \varphi \rangle \cdot \langle f \rangle$. The integral of $p$ over the top plate yields the top plate force, $F$, which reflects the mean field relation for $p$. When the top plate barely touches the grains, $F$ should be null and, due to the density matching, so should $\langle f \rangle$. We subtract the small minimal measured values from $\langle f \rangle$ and $F$, which are present at the instant the plate touches the grains, and which we attribute to measurement inaccuracies. This guarantees that *F*=0 when ⟨*f*⟩=0. The coefficient of proportionality is the averaging volume, *V*, divided by the top plate area. As shown in Fig.3c, this relation is remarkably well respected.

**MM4: Experimental Methods**

The granular material used here consists of 514 hydrogel beads [17,20]. The beads are approximately spherical and roughly monodisperse with a typical diameter of 2.1 ± 0.1cm. They are immersed in a water-polyvinylpyrrolidone 360,000 MW(PVP) solution in order to match the particle index of refraction to the surrounding fluid. This matching allows interior optical access which is necessary for the refractive index matched tomography technique used here [16]. Since the particles are almost entirely composed of water, they are also nearly density matched with the fluid; the particle density is less than 10kg/m³ greater than the fluid density. By itself, the fluid-particle system is completely transparent. In order to obtain contrast, we dye the particles with a hydrophobic fluorescent dye (Nile Blue 690) which can be excited with a laser sheet, which we scan across the system. This laser sheet (Lasiris SNF 635nm, 25mW) moves on a fast linear stage, which sequentially illuminates the sample slice-by-slice. A fast camera (Basler ava1000-120) equipped with lens and long pass filter records the fluorescent image of each slice. This produces a sequential sequence of particle cross sections. The camera is mounted on the same stage as the laser, as in the design of Ref [24]. A complete scan typically consists of 360 slices. We confine the granular system in a rectangular box with base size 16.5cm×16.5cm and carry out cyclical uniaxial compression/decompression using a stage-controlled piston as described in the main text. The top piston is made from a 6 mm thick perforated sheet, which allows flow of index-matching fluid into and out of the packing. Except for gravitational gradients, the pore pressure inside the fluid is uniform and matched to the ambient atmosphere. The piston is driven by a linear stage (Newport MTM250) and controller (Newport XS4), with a step resolution of 1 micrometer. The force on the piston plate is measured with a force sensor (Loadstar RSB4-005M-A). The compression speed during compression/decompression is 0.1mm per second to reduce fluid induced shear stresses on the particles and to drive the granular system completely quasi-statically. The spacing between the piston plate and the walls is small enough that particles cannot escape confinement, so the number of particles during an experiment is constant. The container height depends on the compression level, but a typical uncompressed height is about 15cm. More experimental details can be found elsewhere [18].